# Berta Karlik – The *Grande Dame* of the Vienna Radium Institute

Brigitte Strohmaier, University of Vienna, Faculty of Physics – Nuclear Physics, Vienna, Austria

## Abstract

Berta Karlik (1904–1990) was an Austrian physicist who was not only among the early radioactivity researchers and nuclear physicists in Vienna, but also pioneered a woman's academic career in Austria: She was the first woman at the University of Vienna to acquire the *venia legendi* (right of teaching at a university) in physics, and the first full professor at a philosophical faculty in Austria. For almost thirty years she was the head of the Institute for Radium Research of the Austrian Academy of Sciences.

This article is a translation of Brigitte Strohmaier, Berta Karlik – Die *Grande Dame* des Wiener Radiuminstituts, Mensch – Wissenschaft – Magie, Mitt. Österr. Ges. für Wissenschaftsgesch. 27 (2010) 91–108.

## Keywords

## 1. Family and education

Karlik was born on January 24, 1904, to the lawyer Dr. Carl Karlik, director of the mortgage company of the province of Lower Austria, and his wife Karoline, née Baier, as eldest of three siblings in Mauer near Vienna.[1] At first she received private tutoring by her mother, later on, she attended primary school in Mauer and then high school in Wenzgasse 7 (Vienna's 13th district).

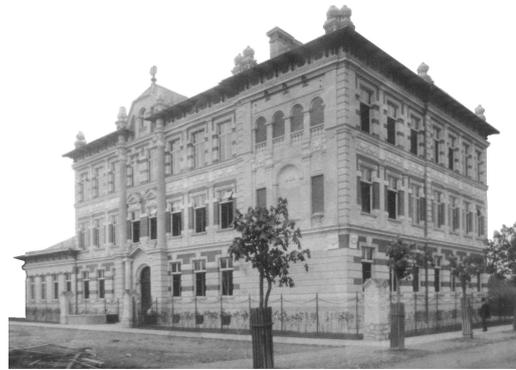

Fig. 1: The *Mädchenlyzeum* in Wenzgasse 7 in 1906 (Courtesy GRG 13 Wenzgasse)

This school had been founded in 1904 as *Mädchenlyzeum*, a type of girls' school that was meant to train the girls for employment. Later on, the *Reformrealgymnasium* was created which added to the *Lyzeum* the upper grades of a *Gymnasium* (high school), offering the students the general certificate of education (*Matura*) and hence entitling them to enroll at any university without further examinations.[2] In 1920, the first *Matura* took place in the Wenzgasse school. Karlik obtained hers with distinction in 1923. In fall of 1923, Berta Karlik enrolled at the University of Vienna as regular student of physics, mathematics, and chemistry. She was particularly interested in radioactivity research; therefore, she worked her doctoral thesis at the Institute for Radium Research.

Radioactivity consists in the spontaneous emission of radiation and was discovered by Henri Becquerel in Paris in 1896. The phenomenon was further explored by Marie and Pierre Curie who succeeded in isolating two radioactive elements, radium and polonium, in 1898.

In Vienna, scientists had begun investigating radioactivity at the Physics Institute of the University as early as 1899, but did not have adequate laboratories and instruments at their disposal. This motivated Dr. Karl Kupelwieser, a lawyer, to donate part of his property (500,000 crowns, about two million dollars) for the construction and equipping of a building which should be owned by the Imperial Academy of Sciences and serve for the (physical) investigation of radium. This "Institute for Radium Research" was located in Boltzmanngasse 3 (at that time Waisenhausgasse; ninth district) and dedicated in 1910. It was the first "Radium Institute" in the world and example for similar institutions built later on in Paris, St. Petersburg, and Prague. Already in the first few years of its existence, scientists succeeded in great achievements, e. g., the discovery of cosmic rays and the development of the method of radioactive indicators (tracers) which were awarded the Nobel Prize in Physics and Chemistry, respectively.

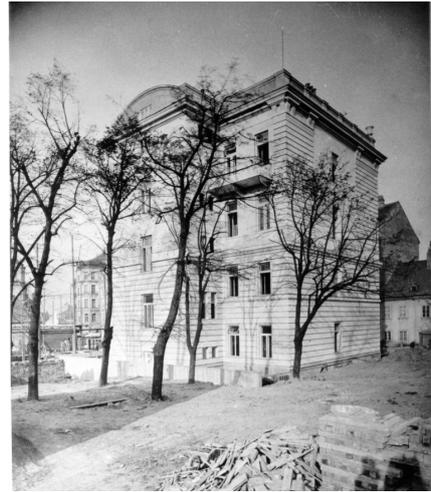

Fig. 2: The Vienna Radium Institute in 1910, view from the rear courtyard (Austr. Acad. Sci., Photo Archive A-0142-C)

Although the Institute for Radium Research belonged to the Academy of Sciences according to Kupelwieser's donation, its statute provided that it was operated by the government and part of the physicists doing research there were employed by the Ministry of Education, i. e., the University of Vienna, and hence fulfilled teaching obligations.[3]

Karlik's Ph.D. graduation took place in March 1928. Thereafter, she completed her studies for the teaching profession at high schools, passing her finals in fall of 1928.[4] In the school year 1928/29 she served as probationer at the girls' high school (*Mädchenrealgymnasium*) in Albertgasse in Vienna's eighth district. In the summer semester of 1929, she acted as substitute for a teacher on sick leave. Her teaching was considered particularly successful and earned her unusual popularity. Undecided whether she should practise the profession of a teacher or a scientist, she returned to the Radium Institute and worked there for no pay, as many other women did in these early years of nuclear physics. Like most of these women – the portion of females was about one third – Karlik was a member of the *Verband der Akademikerinnen* (VAÖ), the Austrian branch of the International Federation of University Women. The VAÖ granted her a fellowship which enabled her to spend one year in England in 1930/31.[5] During that time, she stayed at Crosby Hall, a building of the British University Women in London, and worked for the major part of this year on crystallographic investigations at the Royal Institution in London with Sir William Bragg. Part of the time she spent in Cambridge and familiarized with Lord Rutherford's nuclear-physics research. On top of all this, she engaged in medical applications of radium, a field in which she was extremely interested all her life. In this connection, she visited the physical departments of leading English hospitals. In addition, she travelled from London to the Paris Radium Institute and made the acquaintance of Mme. Curie.[6] The research stays in England and Paris were for Karlik an extraordinarily fascinating experience which led to her decision to pursue a career as scientist rather than a high-school teacher.

## 2. Karlik's university career

Karlik returned to the Vienna Radium Institute in 1932 and at first did unpaid work again. In the following year she was employed as scientific assistant, though. In 1937, she acquired her *Habilitation* (right of teaching at universities) on the basis of the work "Detection Limits of the Heavier Noble Gases in Helium" *(Die Grenzen der Nachweisbarkeit der schwereren*

*Edelgase in Helium*). She started a very intense and thematically diversified teaching activity: She gave lectures in physics of noble gases, use of radium in medicine, spectral analysis, atomic constants, electron diffraction, as well as radioactivity and nuclear physics.[6]

After Austria's annexation to Nazi Germany in 1938, the universities were subject to the Reich Ministry for Education, Science and Adult Education, and the German university organisation was introduced in what was now called *Ostmark*, too. When the "non-Aryan" staff members of the Radium Institute were removed, this was detrimental to the Institute and for Karlik meant a severe loss of colleagues with whom she had collaborated.

Karlik was appointed *Dozent neuer Ordnung* (university lecturer new type) according to her application in 1939. After the war, directorship of the Radium Institute was provisionally transferred to her. In conjunction, she obtained the title (not position) of an associate university professor. In 1947, she became the permanent director of the Radium Institute.[1] In 1950, the Federal Ministry for Education altered the dedication of an existing associate professorship to nuclear physics and Karlik was appointed associate professor. In 1956, this position was changed to full professorship.[6] At that time, science was part of the philosophical faculty. Karlik was, therefore, the first female full professor at a philosophical faculty of an Austrian university.

In September 1974, she became *professor emerita* which ended her active service at the University of Vienna. She was succeeded as director of the Radium Institute of both the Austrian Academy of Sciences and the University of Vienna by Herbert Vonach.

## 3. Karlik's scientific work
### 3.1. Scintillations

Karlik worked on her doctoral thesis at the Vienna Radium Institute as of 1926; head of the Institute was Stefan Meyer. At that time, the Institute's work focused on the study of "atomic disintegrations" (*Atomzertrümmerungen*) under the guidance of the Swedish guest scientist Hans Pettersson: One had to direct α-particles of radioactive decay to stable atomic nuclei and detect the emitted particles. Such an artificial nuclear reaction had first been described by Ernest Rutherford in England in 1919, namely the emission of protons following the incidence of α-particles on nitrogen gas.

For registering the emitted particles, primarily protons, one used the scintillation method which relies on the fact that these release tiny light flashes when incident on certain materials, in particular sphalerite (zinc blende, ZnS) screens. The weak luminosity had to be observed and counted in dark chambers under microscopes or magnifying glasses, a way of particle detection which was tedious and subject to error. In fact, the nuclear-reaction results obtained by Pettersson's group at the Vienna Radium Institute began deviating significantly from those of Rutherford's team in Cambridge.[7] This was the incentive for efforts at the Radium Institute, on one hand to develop alternative detection techniques, on the other, to improve the scintillation method in order to make it objective and reproducible to a higher extent. Therefore, Berta Karlik investigated in her thesis the dependence of scintillations on the quality of zinc sulfide and the nature of the scintillation process.[8] She experimented with α-particles and by photometric means determined the relationship between their energy and the brightness of scintillations. The latter is proportional to the ionization caused in the sphalerite crystal, as Karlik found out.

At the end of 1927 the Cambridge group sent a commission to Vienna. The results obtained by Pettersson's collaborators by visual scintillation counting did not withstand the examination. Based on Karlik's findings, observing scintillations with the human eye could now be abandoned in favor of registering the scintillation light by means of a photoelectric cell.[9] In this way, the (objective) measurement of the photo current replaced the (subjective) visual recognition and counting of light flashes.

In 1933 Karlik received the Haitinger Prize for physics of the Academy of Sciences for her work in the field of luminescence.[10]

### 3.2. Crystal structure

When Karlik did research in London with William Bragg (1931), she and two female colleagues concentrated on the structure of cubic crystals of pure elements and compounds. Their work resulted in a tabular list, consisting of a compilation of original papers in which details of the structure of cubic crystals were determined and stated. These tables were published in the book "Crystal Structure of Elements and Compounds"[11] and was meant as a tool for those who applied x-ray structure analysis in practice as well as those who used it in pure-science research: The collection of the most important data of the large number of cubic crystals should make it possible to find the most recent information pertaining to a certain substance.

### 3.3. Fluorescence of fluorite

Karlik investigated the excitation of luminescence of various colors observed in certain minerals, the fluorites, on exposure to ultraviolet light. In fact, the term fluorescence for this phenomenon was derived from the name fluorite. The light emission occurs at luminosity centers which can be removed by heating and regenerated by radioactive irradiation.

Collaborating with colleagues, in particular Karl Przibram who was an expert in the field of fluorescence, Karlik between 1933 and 1937 dealt with the question whether impurity atoms in the crystal lattice of fluorite (chemically calcium fluoride) play a role in fluorescence phenomena. It turned out that small additions of rare-earth elements are responsible for the observed luminescence in the red, blue and yellowish-green region,[12] whereas a weak bluish-green fluorescence is traced back to the presence of uranium compounds.[13]

### 3.4. Uranium in seawater

In the 1930s, concentrations of various substances in the sea were determined quantitatively all over the world, for reasons of biological, geochemical or industrial interest. Hans Pettersson who was an oceanographer in Göteborg, but had influenced the research program of the Vienna Radium Institute already since 1923, turned the attention on the study of radioactive elements in sea water. The concentrations of uranium, radium, and thorium were to be determined separately. Samples were taken at the West coast of Sweden at first, later at different places of the oceans.

The task of measuring the uranium content was entrusted on Berta Karlik and Friedrich Hernegger, radiochemist at the Radium Institute. He had developed a method to determine smallest amounts of uranium.[14] For sample preparation Karlik travelled to Sweden several times in the 1930s; the measurements themselves were performed at the Radium Institute.

The significance of the investigation consisted in the combined results for uranium, thorium, and radium content. As in the natural decay chains the radioactive decay of uranium produces thorium and that of thorium radium, expectations regarding the amounts of thorium and radium were derived from the measured uranium concentration. The measured content of Th and Ra was much lower than had been calculated, though, which was explained by chemical processes in sea water: Uranium dropping from the water to the bottom of the sea causes its progeny to be reduced in sea water, whereas enriched in deep-sea sediments.[15]

## 3.5. Discovery of the natural occurrence of element 85

It was one desire of nuclear physics in the early 1940s to fill the remaining gaps in the periodic table of elements. As radioactive elements were mainly identified by means of the characteristic energies of the particles they emit, search for new elements meant search for new radiations.

Together with Traude Bernert who came to the Vienna Radium Institute as volunteer in 1942, Karlik investigated whether dual (α and β) decay known from some nuclei of the three radioactive chains might also be present in other members of the decay chains. This was related to the question whether the elements with atomic numbers 85 and 87 occurred in nature as these would be produced by such transitions not yet observed.

Element 85 appears in the natural decay chains as daughter product of three polonium isotopes after β-decay with extremely small branching ratios. Karlik and Bernert were able to detect the radiations of formation and decay of that isotope (with mass number 218) of element 85 which belongs to the uranium-radium series.[16]

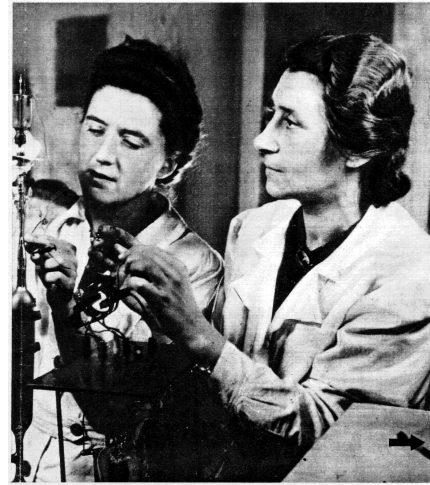

This work is often praised as the greatest success in Karlik's scientific work as it amounts to the discovery of an element. However, physicists in the U.S.A. had synthesized the element two years earlier by means of artificial nuclear reactions in the laboratory. As synthesizing an element was rated as discovery, Karlik and Bernert were not acknowledged as discoverers of element 85 altogether, but of its natural occurrence only. Therefore, they were not credited with its naming. They would have chosen the name Viennium; the U.S. discoverers named it astatine.

The Austrian Academy of Sciences granted Karlik and Bernert the Haitinger Prize in 1947 for their work on element 85.[17]

Fig. 3: Bernert and Karlik, 1942 (Archive R. & L. Sexl)

## 4. Karlik's activities in the Nazi era

Stefan Meyer and many of the collaborators (of either gender) of the Radium Institute had to leave the Institute after Austria's annexation to Hitler Germany. The Institute became part of the Four-Year-Plan Institute for Neutron Physics, financed by the Reich Office for Economic Development (*Reichsamt für Wirtschaftsausbau*) in Berlin. Under its new head (Gustav Ortner) the Institute's research concentrated on fields pertinent to war interests.

In 1938 the VAÖ was dissolved, as it was part of an international organization. In this situation, Karlik made use of her numerous contacts to university women all over the world to help threatened colleagues find jobs abroad and a way to emigrate. All steps necessary to save the threatened persons could be taken outside the German Reich only; visits with Hans Pettersson in Göteborg and on a measuring station at the Swedish West coast in connection with the determination of uranium in sea water offered her the possibility to do so.[18]

Emigrated colleagues commented on Karlik's engagement full of appreciation already during the war: Among the "entirely decent" one named "most of all Berta Karlik, whose character is well known and for whom one has to be afraid because she becomes so personally involved."[19] In post-war years it read: "Berta Karlik's fearless conduct, her constant helpfulness, and loving devotion for all in need, became particularly prominent in this time."[20]

## 5. Science management – The Radium Institute 1945–1974

After the end of World War II Stefan Meyer was re-instated as head of the Institute, but did not exert his function any more due to his advanced age. The Academy of Sciences entrusted Berta Karlik with provisional directorship.[21]

Before one could even think of scientific work, comprehensive salvage work was required. In April 1945 the battles in Vienna had taken place also in immediate neighborhood of the Radium Institute for several days. The adjacent buildings of the physical and chemical institutes had been hit by some smaller artillery strikes, and even though the Radium Institute had not been hit, it was in a deplorable state. The precious radium preparations had been brought to a shelter beneath the *Hofburg* (Imperial Castle) when the bombardment of Vienna began, and were later transferred to a salt mine in Hallein. The instruments had been brought partially to Schwallenbach in Wachau, partially – in view of the advancing Red Army – further to the West, to Thumersbach near Zell am See. The library had been packed and saved in the cellar of the Institute, where also the few apparatus still in operation were carried at each air raid. After the battles had ended, Karlik and some of her colleagues started clearing the Institute. In 1950 Karlik wrote retrospectively:[21]

> The Institute gave a sad sight. The empty rooms were covered with pieces of broken glass, debris, and dust which had been slung in by bomb hits in the surroundings…The following weeks were dedicated to cleaning and clearing, the library was put in its place again, the remaining apparatus collected from the cellars…Life was indescribably hard, a long march brought us to the Institute, at noon a few potatoes were boiled, at first one could not even think of scientific work. Apart from the lacking apparatus, the usage of electricity and gas was extremely restricted. The heating conditions in the first and second post-war winter were catastrophic, so that the Institute's operation had to be reduced to a minimum in the cold season.

Still in 1945, Karlik succeeded with the difficult endeavor to bring home the Institute's preparations from the regions occupied by foreign (allied) military. At the end of September 1945 she had been informed that U.S. officials had found a larger amount of radium in the salt mine of Hallein which they kept safe in Salzburg. She could prove that these were the stocks of the Institute, traveled over the demarcation line (separating Soviet and Western occupation zones) to Salzburg with Friedrich Hernegger and received the radium. A U.S. military truck was equipped with a brick wall according to Karlik's calculations in order to protect the driver from radiation, and accompanied by an armed jeep they drove to Vienna. In the following months also the evacuated instruments could be brought back.

After the Austrian Academy of Sciences had entrusted Berta Karlik with permanent directorship of the Institute for Radium Research in 1947, she worked out a plan for the Institute's future scientific activities. It comprised mainly the construction of a particle accelerator (for deuterons) to generate monoenergetic fast neutrons, and the installation of a facility for dating with the radiocarbon method.

## 5.1. The neutron generator

Already in 1948 one began with the construction of a neutron generator, a facility which by means of a nuclear reaction produces fast, i.e., energetic neutrons.[22] The study of nuclear reactions which were in turn induced by these neutrons on various materials became the main field of the Institute's scientific research. Its name was, therefore, extended to be *Institut für Radiumforschung und Kernphysik* (IRK, Institute for Radium Research and Nuclear Physics) in early 1955. The knowledge of these nuclear reactions is required for applications in technology, in particular for the design of nuclear power plants, and in medicine for the production of radioisotopes to be used for diagnosis and therapy, as well as in pure science to understand the reaction mechanisms and describe them by theoretical models. It turned out that the neutron generator, together with the associated detector and data-acquisition systems, needed increased space and at the same time radiation shielding which was prescribed for safety reasons by law. At first, a new building for this facility was under consideration, but no appropriate place existed near the Radium Institute, and a branch at larger distance would have been disadvantageous to the teaching operations. Karlik solved the problem by adding one floor to the institute building in 1965 which could be paid for with residual money from the Kupelwieser donation.[23] The additional floor accommodated a shielded accelerator and experiment room (a concrete bunker with a mass of over 700 tons[24]) as well as control and other rooms.

For modernizing the Radium Institute Karlik was honored with (one half of) the Schrödinger Prize of the Austrian Academy of Sciences in 1967.[25]

## 5.2. The facility for radiocarbon dating

A further innovative field in the Institute's scientific work was dating by means of the $^{14}$C (radiocarbon) method.[26]

The radionuclide $^{14}$C is formed by cosmic rays in the atmosphere and via the compound $CO_2$ enters tissue of plants, animals and humans. As long as the tissue is alive the contained $^{14}$C is in equilibrium with that of the environment. After the death of the living creature the amount of $^{14}$C in its tissue decreases with a half-life of 5730 years so that the $^{14}$C content is correlated to the time since death, i.e., the age of the sample. The $^{14}$C content is determined through measurement of a very weak β radiation low in energy.

The radiocarbon method had been developed by Willard F. Libby in the U.S.A. in 1949/50. The setup of the corresponding facility for $^{14}$C dating at the Vienna Radium Institute was started still in the 1950s. Routine dating of sample material from archeology, early history, geography and geology, glaciology, climatology and other fields was performed as of 1962.

## 5.3. The isotope office

In 1949 Karlik installed the so called *Isotopenstelle* (isotope office)[27] at the Radium Institute which was in charge of the import of artificially radioactive substances from England, later on also from the U.S., and which passed on the preparations to the purchasers for use in research, technology and medicine. Karlik's efforts to go for such an office at the Radium Institute were motivated by the fact that at that time no legal regulations for shipment and inventory management of radioactive substances existed. Therefore, it made sense that import and distribution be taken care of by a central agency which was competent of handling these preparations and recorded the stocks of imported radioisotopes. At the same time, the isotope office was of advantage to the Radium Institute in that at the beginning it bore part of the salary of employees, widened their experience and occasionally enabled the use of the preparations passing the Institute as well as the detectors and radiation-protection devices which had to be bought for the isotope office for the Institute's purposes.

# 6. Karlik's functions in national and international boards
## 6.1. Standards and metrology

The Institute for Radium Research was obliged to measure and gauge radioactive preparations according to its statute. The International Radium Standard Commission founded in 1910 in which Stefan Meyer had functioned first as secretary, later as president, was followed in 1949 by the Joint Commission on Standards, Units and Constants of the international organizations of pure- and applied-science oriented physics and chemistry. When in 1950 – after Meyer's death – Berta Karlik was elected Advisory Councilor of this board, she felt satisfaction that Austria was present again in the authoritative international radium commission, and found herself once more in the situation of continuing a tradition in which Meyer had preceded her.[28] This commission was responsible for international agreements regarding the use of symbols, units, nomenclature and standards.

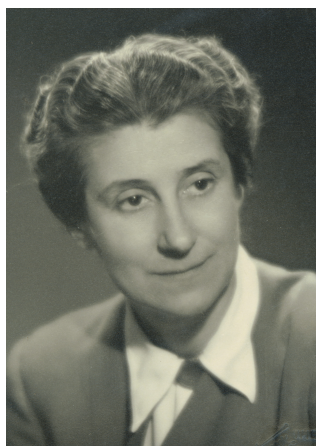

Fig. 4: Berta Karlik, 1951 (Austr. Acad. Sci., Photo Archive P-1357-B)

In 1957 the Joint Commission on Standards, Units and Constants was dissolved and its agenda was taken over by the *Bureau International des Poids et Mesures*. Karlik was member of its Consulting Committee for Ionizing Radiations (*Comité Consultatif pour les Radiations Ionisantes*).[29]

## 6.2. Radiation protection

In view of a legal regulation of handling radiation risks the Austrian Federal Ministry for Social Administration presented the first draft of a corresponding law in 1958. Karlik was a member of both the newly installed Radiation Protection Commission of this ministry and the equally newly founded *Zentralstelle für Strahlenschutz der Österreichischen Akademie der Wissenschaften* (Central Agency for Radiation Protection of the Austrian Academy of Sciences) for which she expressed her view about the draft. The radiation protection law was decreed in 1969, the implementation order was issued in 1971.

### 6.3. Nuclear-physics concerns beyond the Radium Institute

In 1951 Karlik was on her own initiative included to the planning of the European nuclear-research center CERN where she put forth efforts to achieve Austria's membership.[30] In addition, she was a member of the advisory commission of the government with regard to the peaceful use of atomic energy. In this function she contributed to establishing the Austrian Atomic Energy Commission and to the foundation of two research reactors (Seibersdorf, Atomic Institute of Austrian Universities), as well as the creation of the International Atomic Energy Agency.

### 6.4. Cooperation in societies and federations

After World War II the Austrian physicists expressed the wish for an autonomous Austrian union, whereas before 1938 a branch of the German Physical Society had represented the professional interests of the Austrian colleagues as well. On a meeting in Graz at the end of 1950 Austrian physicists decided to found the *Österreichische Physikalische Gesellschaft* (ÖPG, Austrian Physical Society). Its goals were defined to be furthering and disseminating physics science in research and teaching. On the first annual general meeting in October 1951 the first board was elected. Karlik participated in both the decision of foundation of ÖPG in 1950 and the tasks of the board to which she belonged until 1954.

In 1951/52 Karlik was president of the Chemical-Physical Society which has existed since 1869.

After the war Karlik also became involved in the reestablishment of the association of Austrian university women (VAÖ). After three years of preparatory work by Karlik and two other women, VAÖ was founded anew in 1948 and held its first general meeting. Karlik was president of VAÖ until 1954.[5]

## 7. Karlik as *professor emerita* – The Radium Institute after 1974

After Karlik had been given emeritus status in 1974 she was succeeded by Herbert Vonach as head of the Institute and full professor for nuclear physics. Being a member of the Austrian Academy of Sciences and, in particular, of the Board of Trustees of the Institute of Radium Research and Nuclear Physics, Karlik continued having determining influence on the Institute's fate. Various problems were treated with her cooperation.

### 7.1. Decontamination of the building Boltzmanngasse 3

Karlik had begun to refurbish the Institute in accordance with the radiation protection law as soon as the implementation order had been out. In 1972, she had nine tons[24] of material with low radium content transported away by the Austrian Armed Forces for storage in the reactor center Seibersdorf. This material consisted of side products and remains of the extraction of radium chloride from tailings the Academy of Sciences had bought from the uranium mine in Joachimsthal in Bohemia; the radium chloride had been given the Radium Institute in 1910.

On Vonach's initiative a large amount of radioactive substances was transported for storage to Seibersdorf again in 1975, namely the exactly gauged standard preparations which had been made from the above radium chloride. Added were radioactive fractions from the preparation of radium in purest state in connection with the atomic weight determination in 1911, and various fractions from the development of a process for radium production from Congolese uranium ores of the *Union Minière du Haut Katanga* in the 1920s. Also contaminated instruments and laboratory vessels were brought to Seibersdorf for legally compliant storage.[31]

The major part of the highest-purity radium chloride which once had been used for atomic weight determinations and for making standard preparations was donated to the Institute for Radiochemistry of the Technical University of Munich for scientific usage.

## 7.2. Science-history work

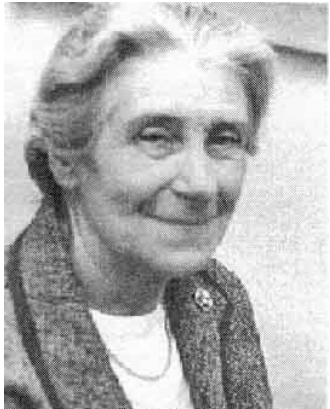

Karlik authored a number of science-history writings[32] among which the obituaries[33] are particularly prominent as they reveal her profound physics expertise and knowledge of human nature as well as her affectionate wording. Moreover, she documented and compiled the scientific correspondence of several physicists.[34]

Together with Erich Schmid she published the book *Franz Serafin Exner und sein Kreis* (Franz Serafin Exner and his circle) in 1982 which describes the life and environment of this famous physicist and first director of the Radium Institute.[35]

Fig. 5: Berta Karlik around 1984 (Austr. Central Library for Physics)

## 7.3. The end of the Institute for Radium Research and Nuclear Physics of the Austrian Academy of Sciences

The university-organization law of 1975 was implemented starting in 1976. Each institute had to declare the extent of representation of its subject, its designation and faculty affiliation. This was often interpreted as re-foundation of the institutes. In the case of the Institute for Radium Research and Nuclear Physics (IRK) the question arose whether the establishment as university institute should be applied for. Karlik rejected this, reasoning that the Institute for Radium Research had been founded in 1910 solely as institute of the Academy of Sciences and, therefore, its existence as university institute was neither given nor necessary.[36] Vonach, the head of the Institute, on the other hand, held the opinion that after all the decades the IRK was certainly a university institute as well, the more so as he had been appointed to a professorship at the Institute for Radium Research and Nuclear Physics of the University of Vienna. Equally, the major part of the staff was employed with the university – with the IRK as place of work – and last, but not least, the receipt of a regular budget was an indication that an Institute for Radium Research and Nuclear Physics of the University of Vienna existed.

Convinced that the foundation of the IRK as university institute according to the university-organization law of 1975 was required to maintain the status of the IRK as academy and university institute, Vonach filed such an application.[37] This step was considered as expressing preference for belonging to the university by some members of the Austrian Academy of Sciences and among these caused a sentiment directed against the IRK. Eventually, this had the consequence that the Institute for Radium Research and Nuclear Physics of the Austrian Academy of Sciences was converted into an Institute for Medium-Energy Physics of the Austrian Academy of Sciences as of January 1, 1987.[38] Directorship of this new institute was transferred to one of Vonach's former collaborators. The Institute for Radium Research and Nuclear Physics from then on existed as a university institute only.[39]

Berta Karlik died on February 4, 1990, shortly after her 86th birthday.[40]

---

[1] Berta Karlik, *Curriculum vitae*, Archive Radiumforschung, Archive Austr. Acad. Sci.

[2] 100 Jahre Wenzgasse. Commemorative document on the 100th anniversary of GRG 13 (Gymnasium & Realgymnasium 13), Vienna (2004).

[3] Statute of the Institute for Radium Research, Almanach Kaiserl. Akad. Wiss. 61 (1911) 215.

[4] Commission for teaching examination 18/1928, Berta Karlik, Archive Univ. of Vienna.

[5] Mitt. VAÖ 66, Special issue 3A, Vienna (1997).

[6] Personnel file Berta Karlik (PH PA 2152), Archive Univ. of Vienna.

[7] Karl Przibram, 1920–1938, Festschrift des Institutes für Radiumforschung anlässlich seines 40jährigen Bestandes, Sitzungsber. Österr. Akad. Wiss., math.-naturw. Kl. IIa 159 (1950) 27; Mitt. Inst. Radiumf. 470 (1950).

[8] Berta Karlik, Über die Abhängigkeit der Szintillationen von der Beschaffenheit des Zinksulfids und das Wesen des Szintillationsvorganges, Thesis Univ. of Vienna (1927); also: Sitzungsber. Akad. Wiss. Wien, math.-naturw. Kl. IIa 136 (1927) 531; Mitt. Inst. Radiumf. 209 (1927).

[9] Berta Karlik, Elisabeth Kara-Michailova, Über die durch α-Strahlen erregte Lumineszenz und deren Zusammenhang mit der Teilchenenergie, Sitzungsber. Akad. Wiss. Wien, math.-naturw. Kl. IIa 137 (1928) 363; Mitt. Inst. Radiumf. 222 (1928).

[10] Prize Awards Acad. Sci. 1933, Almanach Akad. Wiss. Wien 83 (1933) 151.

[11] Berta Karlik, I. E. Knaggs, Tables of cubic crystal structures, in: Cubic Crystal Structure of Elements and Compounds, with a section on alloys by C. F. Elam, Adam Hilger Ltd., London (1932).

[12] Herbert Haberlandt, Berta Karlik, Karl Przibram, Synthese der blauen Fluoritfluoreszenz, Anz. Akad. Wiss., math.-naturw. Kl. 70 (1933) 301; Mitt. Inst. Radiumf. 325a (1933); iidem, Synthese der grünen Tieftemperaturfluoreszenz des Fluorits, Anz. Akad. Wiss., math.-naturw. Kl. 71 (1934) 1; Mitt. Inst. Radiumf. 331a (1934); iidem, Zur Fluoreszenz des Fluorits II, Sitzungsber. Akad. Wiss. Wien, math.-naturw. Kl. IIa 143 (1934) 151; Mitt. Inst. Radiumf. 336 (1934); iidem, Zur Fluoreszenz des Fluorits III. Das Linienfluoreszenzspektrum, Sitzungsber. Akad. Wiss. Wien, math.-naturw. Kl. IIa 144 (1935) 77; Mitt. Inst. Radiumf. 352 (1935).

[13] Herbert Haberlandt, Berta Karlik, Karl Przibram, Zur Fluoreszenz des Fluorits IV. Über einen Urannachweis in Fluoriten und über die Tieftemperaturfluoreszenz, Sitzungsber. Akad. Wiss. Wien, math.-naturw. Kl. IIa 144 (1935) 135; Mitt. Inst. Radiumf. 354 (1935).

[14] Friedrich Hernegger, Methoden für einen empfindlichen Urannachweis in Quellwässern und Quellsedimenten, Anz. Akad. Wiss. Wien, math.-naturw. Kl. 70 (1933) 15; Mitt. Inst. Radiumf. 301a (1933).

[15] Berta Karlik, The uranium content of seawater, in: Ernst Föyn, Berta Karlik, Hans Pettersson, Elisabeth Rona (Eds.), The Radioactivity of Seawater, Göteborgs Kungl. Vetenskaps- och Vitterhets Samhälles Handlingar, Femte Följden, Ser. B6, Nr. 12, Meddelanden Oc. Inst. Göteborg 2 (1939) 7.

[16] Berta Karlik, Traude Bernert, Über eine vermutete β-Strahlung des Radium A und die natürliche Existenz des Elementes 85, Naturwiss. 30 (1942) 685; eaedem, Über eine dem Element 85 zugeordnete α-Strahlung, Sitzungsber. Akad. Wiss. Wien, math.-naturw. Kl. IIa 152 (1943) 103; Mitt. Inst. Radiumf. 449 (1943); eaedem, Eine neue natürliche α-Strahlung, Naturwiss. 31 (1943) 298; eaedem, Ein weiterer dualer Zerfall in der Thoriumreihe, Naturwiss. 31 (1943) 492; eaedem, Über zwei neue α-Strahlungen in der Thorium- und in der Actiniumreihe, Anz. Akad. Wiss., math.-naturw. Kl. 81 (1944) 2; Mitt. Inst. Radiumf. 450a (1944).

[17] Prize Awards Austr. Acad. Sci. 1947, Almanach Österr. Akad. Wiss. 97 (1947) 195.

[18] Berta Karlik, Letter to Ellen Gleditsch, Bockholmen, Bornö, September 11, 1938, Archive Radiumforschung, Archive Austr. Acad. Sci.

[19] Marietta Blau, Letter to Fritz Paneth, Oslo, April 23, 1938, Archive Radiumforschung, Archive Austr. Acad. Sci.

[20] Josef Mattauch, Letter to Stefan Meyer, Agra, April 22, 1947, in: Stefan Meyers Memorandum im Kommissionsbericht über die Sitzung des Kuratoriums des Instituts für Radiumforschung am 13. Mai 1947, Archive Radiumforschung, Archive Austr. Acad. Sci.

[21] Berta Karlik, 1938–1950, Festschrift des Institutes für Radiumforschung anlässlich seines 40jährigen Bestandes, Sitzungsber. Österr. Akad. Wiss., math.-naturw. Kl. IIa 159 (1950) 27; Mitt. Inst. Radiumf. 470 (1950).

[22] Traude Matitsch, Rudolf W. Waniek, Hans Warhanek, Die Teilchenbeschleunigungsanlage des Instituts für Radiumforschung, Sitzungsber. Österr. Akad. Wiss., math.-naturw. Kl. IIa 165 (1956) 273; Mitt. Inst. Radiumf. 523 (1956); Heinrich Münzer, Versuche zur Steigerung der Neutronenausbeute an einer Deuteronenbeschleunigungsanlage mit schwerem Eis- bzw. Tritiumtarget, Sitzungsber. Österr. Akad. Wiss., math.-naturw. Kl. IIa 165 (1956) 267; Mitt. Inst. Radiumf. 524 (1956).

[23] Report of the Institute for Radium Research and Nuclear Physics, Annual Session 1966, Almanach Österr. Akad. Wiss. 116 (1966) 227.

[24] 1 ton = 1000 kg.
[25] Erich Schmid, Announcement of Prizes Awarded by the Austrian Academy of Sciences in 1967, Almanach Österr. Akad. Wiss. 117 (1967) 146.
[26] Heinz Felber, Peter Vychytil, Messanordnung für energiearme β-Strahlung geringer Intensität, speziell zur Altersbestimmung nach der Radiokohlenstoffmethode, Sitzungsber. Österr. Akad. Wiss., math.-naturw. Kl. IIa 170 (1961) 179; Mitt. Inst. Radiumf. 546 (1961).
[27] Report of the Institute for Radium Research, Annual Session 1950, Almanach Österr. Akad. Wiss. 100 (1950) 268.
[28] Report of the Institute for Radium Research, Annual Session 1951, Almanach Österr. Akad. Wiss. 101 (1951) 218.
[29] Report of the Institute for Radium Research and Nuclear Physics, Annual Session 1961, Almanach Österr. Akad. Wiss. 111 (1961) 312.
[30] Berta Karlik, Letter to the Board of Trustees of the Institute for Radium Research of the Austrian Academy of Sciences, Vienna, October 28, 1952, Archive Radiumforschung, Archive Austr. Acad. Sci.
[31] Herbert Vonach, Comments on the radioactive substances to be brought to Seibersdorf for storage, Vienna, May 14, 1975, Archive Radiumforschung, Archive Austr. Acad. Sci.
[32] Berta Karlik was member of the Austrian Society for History of Science from 1980 to 1990.
[33] Berta Karlik, Georg von Hevesy (Nachruf), Almanach Österr. Akad. Wiss. 118 (1968) 261; eadem, Hans Pettersson (Nachruf), Almanach Österr. Akad. Wiss. 119 (1969) 303; eadem, Otto Hahn (Nachruf), Almanach Österr. Akad. Wiss. 119 (1969) 266; eadem, Lise Meitner (Nachruf), Almanach Österr. Akad. Wiss. 119 (1969) 345; eadem, Karl Przibram (Nachruf), Almanach Österr. Akad. Wiss. 124 (1974) 379; eadem, Sir George Paget Thomson (Nachruf), Almanach Österr. Akad. Wiss. 126 (1976) 509; eadem, Josef Mattauch (Nachruf), Almanach Österr. Akad. Wiss. 127 (1977) 490.
[34] Berta Karlik, Der wissenschaftliche Briefwechsel von Stefan Meyer, Sitzungsber. Österr. Akad. Wiss., math.-naturw. Kl. II 188 (1979) 219; Mitt. Inst. Radiumf. 717 (1979); eadem, Register des Briefwechsels mit Wissenschaftlern, Sitzungsber. Österr. Akad. Wiss., math.-naturw. Kl. II 195 (1986) 239; Mitt. Inst. Radiumf. 744 (1986); eadem, Register des wissenschaftlichen Briefwechsels von Karl Przibram, Sitzungsber. Österr. Akad. Wiss., math.-naturw. Kl. II 196 (1987) 49; eadem, Register der Briefe von Erwin Schrödinger an Stefan Meyer, Anz. Österr. Akad. Wiss., math.-naturw. Kl. 124 (1987) 71.
[35] Berta Karlik, Erich Schmid, Franz Serafin Exner und sein Kreis, Verlag Österr. Akad. Wiss., Vienna (1982).
[36] Berta Karlik, Documentation of the Status of the Institute for Radium Research and Nuclear Physics between Academy and University, Vienna (1977), Archive Radiumforschung, Archive Austr. Acad. Sci.
[37] Herbert Vonach, Letter to the Federal Ministry for Science and Research, Vienna, November 29, 1976, Archive Radiumforschung, Archive Austr. Acad. Sci.
[38] Sitzung der mathematisch-naturwissenschaftlichen Klasse und Gesamtsitzung der Österreichischen Akademie der Wissenschaften am 11. Oktober 1985; Sitzung der mathematisch-naturwissenschaftlichen Klasse und Gesamtsitzung der Österreichischen Akademie der Wissenschaften am 30. April 1986; Otto Hittmair, Bericht des Sekretärs der math.-naturwiss. Kl. über den Zeitraum 15. Mai 1986–13. Mai 1987, Almanach Österr. Akad. Wiss. 137 (1987) 156.
[39] Note added in 2023: When the IRK of the Austrian Academy of Sciences was converted into a Medium Energy Institute, not only the director's post, but also the appropriations from the Academy of Sciences were transferred to the new institute. Two technicians of the IRK employed with the Academy were committed to the university. A graduate scientist turned down the corresponding offer and left the IRK at all. All these changes caused the IRK university staff to speak of a dissolution of the Institute for Radium Research and Nuclear Physics of the Academy of Sciences and a new foundation of an Institute for Medium-Energy Physics of the Academy of Sciences, rather than a conversion of the former to the latter.
[40] This article summarizes a biographical book (in German) on Berta Karlik which is being prepared.